\documentclass[12pt,twocolumn,a4paper]{article}

\usepackage{times}
\usepackage{graphicx,epsfig}
\usepackage {subfigure}

\title{A relativistic navigation system for space}
\author{Angelo Tartaglia\thanks{E-mail address: angelo.tartaglia@polito.it}, MatteoLuca Ruggiero\thanks{E-mail address: matteo.ruggiero@polito.it}, Emiliano Capolongo\thanks{E-mail address: emiliano.capolongo@polito.it} \\
\small {Dipartimento di Fisica, Politecnico di Torino, corso Duca degli Abruzzi 24, 10129 Torino, Italy,}\\ \small {and INFN, via Pietro Giuria 1, 10126 Torino, Italy} }

\begin{document}
\maketitle

\begin{abstract}
We present here a method for the relativistic positioning in spacetime based on the reception of pulses from sources of electromagnetic signals whose worldline is known. The method is based on the use of a four-dimensional grid covering the whole spacetime and made of the null hypersurfaces representing the propagating pulses. In our first approach to the problem of positioning we consider radio-pulsars at infinity as primary sources of the required signals. The reason is that, besides being very good clocks, pulsars can be considered as being fixed stars for reasonably long times. The positioning is obtained linearizing the worldline of the observer for times of the order of a few periods of the signals. We present an exercise where the use of our method applied to the signals from four real pulsars permits the reconstruction of the motion of the Earth with respect to the fixed stars during three days. The uncertainties and the constraints of the method are discussed and the possibilities of using moving artificial sources carried around by celestial bodies or spacecrafts in the Solar System is also discussed.
\end{abstract}

\section{Introduction}
In ancient times people learnt to travel by sea, far from the cost, looking at the sky. Though they measured time on the basis of the day and night alternation and not much more, Polynesian settlers, using the stars as a guide, were able to sail across the Pacific ocean over thousands of kilometers without getting lost. In the West, once the measurement of time reached modern accuracy and precision with the first marine chronometer, the celestial navigation was the base of the spread of European colonization over the world.
Today the equivalent of the old navigation (and of the present, though by other guidance systems) is represented by the exploration of the Solar System and, possibly, even beyond. Until now however the guidance of the spacecraft is performed from Earth; stars are used at most for trim definition purposes on board.
In fact, the idea of using stars for navigation in space seems sound and appealing, but now it must take into account our better knowledge of the concepts of time and space which General Relativity binds together. In particular, in order to achieve the precision and accuracy needed in space we cannot simply consider the configuration of  ``fixed'' stars and the times at the origin of our travel and at the local position. We need to compare clocks far away in the sky (or at least following known spacetime trajectories or worldlines) with a clock we carry with us.
The idea we are reviewing and presenting in the present paper outlines a fully relativistic navigation system based on the local measurement of the arrival times of electromagnetic signals from sets of at least four pulsating sources, located at known positions in the sky. The first implementation of the idea considers pulsars as sources. However, as we shall see, the same approach can be adopted when the origin of the pulses is an artificial one.

\section{Relativistic positioning}
\begin{figure}[ht]
\centering
\includegraphics[width=1.0\linewidth]{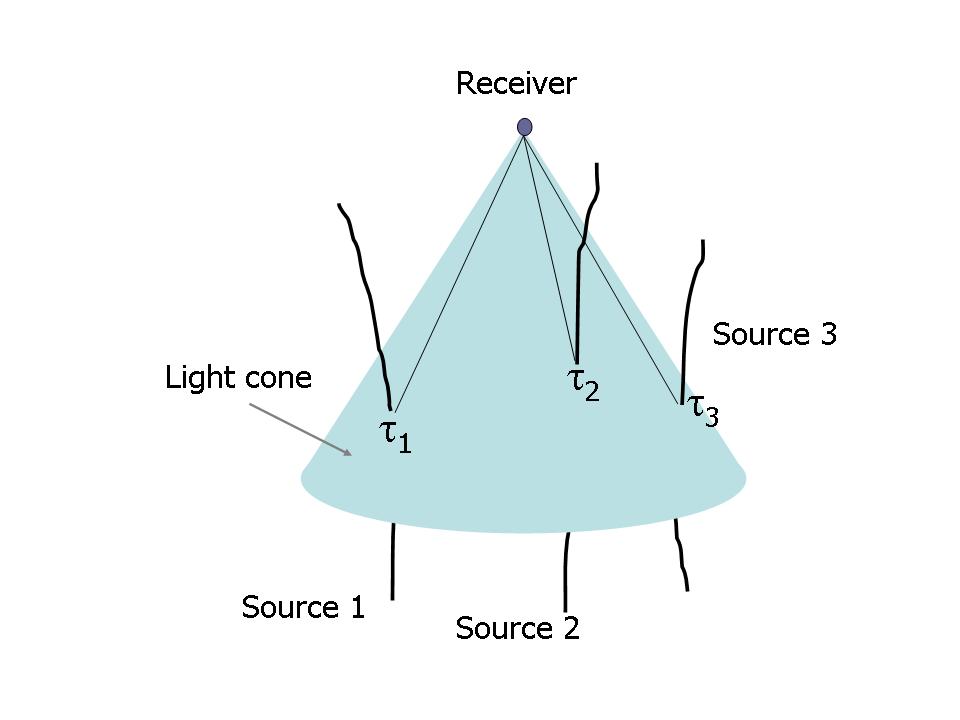}
\caption{View of a pointlike observer with its past light cone; time flows from bottom to top. The black almost vertical lines correspond to the world lines of three light emitters slowly moving around. The straight lines on the surface of the cone are the worldlines of the electromagnetic signals emitted by each source. The numbers $\tau_{1}$, $\tau_{2}$ and $\tau_{3}$ are the proper times at the emission event.}
\label{ff1}
\end{figure}

Any object in spacetime, for instance a pointlike observer,  is represented by a line, actually its worldline, in the four-dimensional continuum. Electromagnetic signals that reach the observer at a given position and time, travel on his past light cone. The situation is schematized in fig. \ref{ff1}.

The figure necessarily represents a three-dimensional spacetime; actually the dimensions should be four, but the geometrical configuration is essentially the same. Imagine you have not less than four broadcasting devices, each one equipped with a clock; their electromagnetic signals can convey the information of the proper time of each emitter at the moment of the emission. The user is reached at any moment by a set of four signals; the information carried by each  signal concerns the identification code of the source and the proper emission time of that signal. The relevant fact is that the set of the four emission times depends on the position of the observer in spacetime. The correspondence between positions and quartets of proper times is one to one, provided the four worldlines of the emitters are linearly independent from one another and as far as we do not consider lensing effects, which are important in strong gravitational fields. Under these conditions we may think to use the four numbers (the four emission times) as good coordinates localizing the receiver both in space and in time (this is the reason why the emitters must be at least four). The basis for this peculiar coordinate set is given by the spacetime trajectories of the four emitters. The four proper times are the \textit{emission coordinates} of the user. This approach has been considered by various authors \cite{Blagojevitc2002,Rovelli2002,Ruggiero2008,Bini2008} and has especially been studied by B. Coll and collaborators \cite{Coll2006,Coll2006a,Coll2006b}.

Our approach is slightly different, but in the end it produces an equally reliable positioning. If we imagine to have a finite source of electromagnetic waves located at space infinity, the wave fronts of the signal will everywhere be planes. Suppose the source emits periodic pulses: geometrically the signal will be a set of non-intersecting planes traveling at the speed of light. If we have four of such pulsed sources, their pulses will fill space with a sort of egg-crate lattice. Each signal from each source may be labeled with a simple ordinal number, so that each cell in the lattice is identified by the labels on one of the corners. The orientation of one family of planes (pulses from one of the sources) is specified when a unit vector perpendicular to one of the planes is given. All this is expressed in three dimensions, but spacetime is four-dimensional, so that the same considerations as above can be made identifying each traveling pulse with a (three-dimensional) hyperplane; its orientation will be given by a four-vector orthogonal to the (family of) hyperplanes. Our four-vector can carry all the relevant information if it is written like this:

\begin{equation}
\chi \doteq \frac{1}{c T}(1,\vec{\mathbf{n}})  \label{eq:deff}
\end{equation}

Here $\vec{\mathbf{n}}$ is a purely spacelike unit three-vector. In an arbitrary Cartesian coordinates system in space its components would be the direction cosines of the vector with respect to the coordinated axes. $T$ is the periodicity of the pulses given in a reference frame where the source is at rest (proper period of the pulses). The factor we put in front of the expression of $\chi$ in (\ref{eq:deff}) produces the same dimensions as for ordinary three-dimensional wave-vectors.
Since we are speaking of electromagnetic pulses, the $\chi$ vector is null or self-orthogonal:

\begin{equation}
\chi \cdot \chi = \frac{1}{c^2 T^2}(1-1)= 0  \label{eq:null}
\end{equation}

 If we consider the covariant version of $\chi$, which is, technically speaking, a 1-form, we know that it has a Hodge conjugate 3-form:

 \begin{equation}
\varpi = \ast\chi  \label{eq:omega}
\end{equation}

The 3-form $\varpi$ is, by construction, perpendicular to the four-vector $\chi$: $\chi \cdot \varpi = 0$. Furthermore, being $\chi$ a null vector, $\varpi$ too is null: $\varpi \cdot \varpi = 0$. In practice, since $\chi$ identifies a direction in spacetime, $\varpi$, as we wrote above, corresponds to a family of three-dimensional hyperplanes perpendicular to $\chi$. When we split the four-dimensional description into space and time ($3+1$ splitting), the projection of the above picture in space gives the familiar view of a set of ordinary bi-dimensional planes (wave fronts) propagating along the direction given by the space components of $\chi$ at the speed of light.

What matters for us is that four families of independent hyperplanes of this sort cover the whole spacetime with a four-dimensional foam of (hyper)cells, each one uniquely labeled by a set of four integers (the ordinal numbers of the hyperplanes, i.e. of individual pulses, meeting at one of the vertices). This configuration permits to position every event in spacetime modulo the edges of a cell. If we identify the sources by the indices $a,b,c,d$ the lengths of the edges will be $cT_a$, $cT_b$, $cT_c$, $cT_d$; each edge is null in the sense of (\ref{eq:null}).

\begin{figure}[ht]
\centering
\includegraphics[scale=0.31]{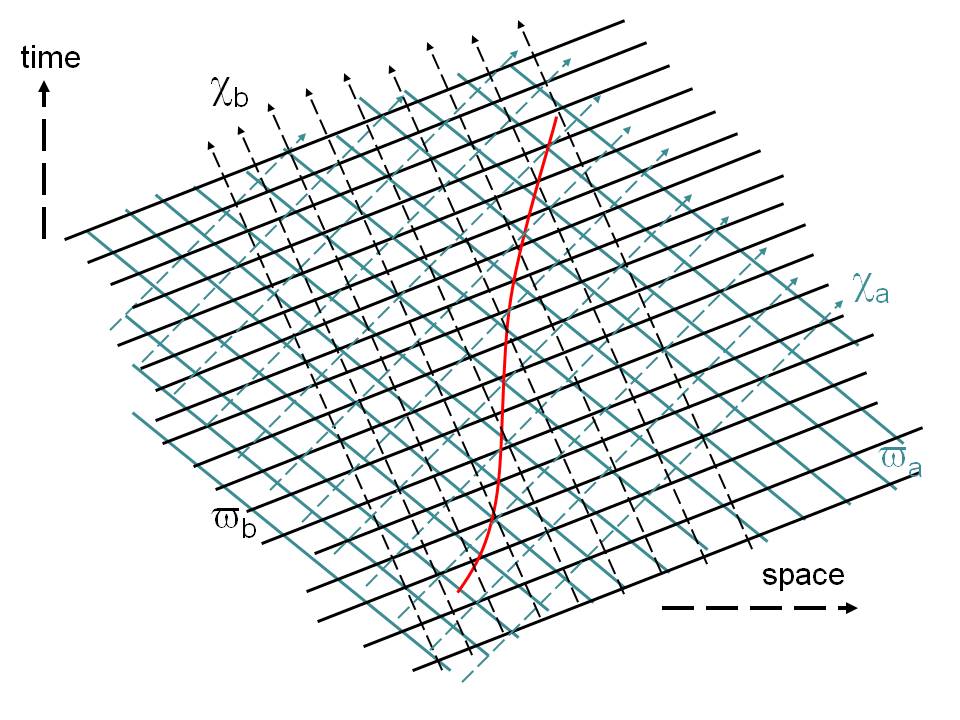}
\caption{Bidimensional view of the positioning pattern described in the text. $\varpi_a$ and $\varpi_b$ identify
two families of hyperplanes; each hyperplane corresponds to a single pulse from a source and may be labeled by
an ordinal number. The dashed lines correspond to the flow lines of the $\chi$ null four-vectors. The continuous wiggling line is the worldline of an user traveling across spacetime. The intersection of the worldline with one
of the hyperplanes identifies the reception event of the corresponding pulse.}
\label{ff2}
\end{figure}

 Now, if we have an observer moving across spacetime, his worldline successively intersects the cells of the ``foam'' we mentioned above. The situation is sketched in fig. \ref{ff2}.

The intersection of the worldline with one of the hyperplanes identifies the reception event of the corresponding pulse. If the receiver is equipped with a clock he can both count the subsequent arrivals of the pulses and measure the proper time intervals between the arrivals, represented in fig. \ref{ff2} by the length of the worldline between two intersections.

If we consider, for instance, an arrival event from source $a$, we may label it with the integer $n_a$. Of course, from the viewpoint of the signals from $b$, the $n_a$ event will have a label somewhere in between an $n_b$ and $n_b + 1$; the same will in general be the case for sources $c$ and $d$. In practice, we may think to use the four numbers $\{n_a;n_b+x_b;n_c+x_c;n_d+x_d\}$ as coordinates to localize the reception event of the $n_a$-th pulse from source $a$. The $x$'s we have introduced are in general $0\leq x < 1$.

\subsection{Linearization}

\begin{figure}[ht]
\centering
\includegraphics[width=1.0\linewidth]{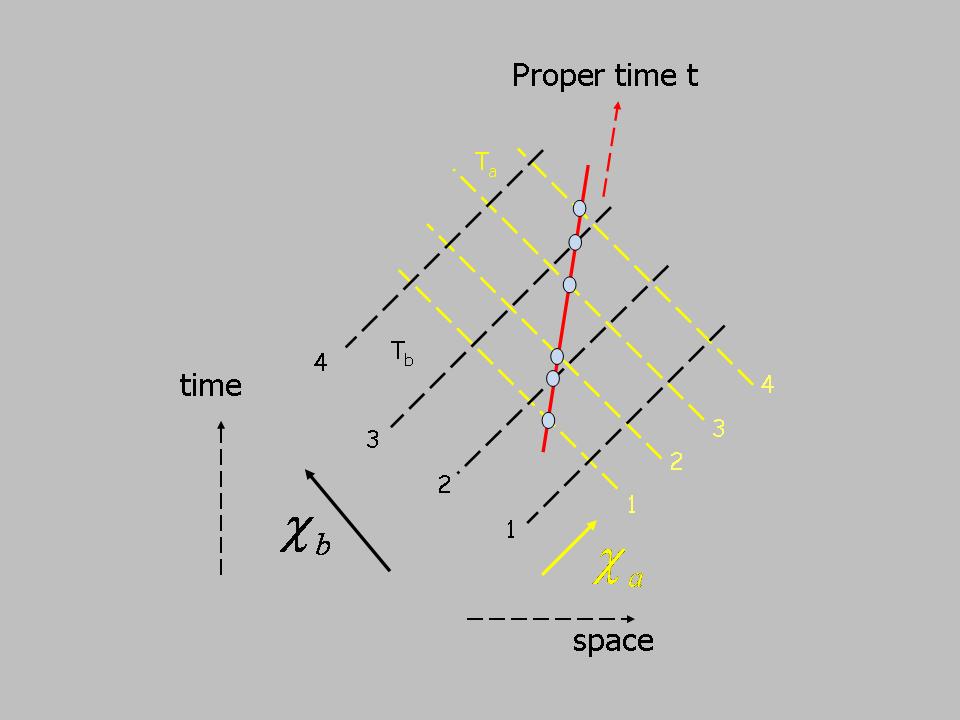}
\caption{Bidimensional representation of the reception of a number of successive pulses arriving from sources $a$ and $b$ according to the scheme of fig. \ref{ff2}; the little ovals identify the arrival events.
When the worldline of the user is straight, or may be thought to be approximately straight, the knowledge of
the time span between the arrivals of the pulses is sufficient to uniquely localize the reception events,
and then the position of the user.}
\label{ff3}
\end{figure}

If we want the above defined coordinates to be useful we must find a practical way to evaluate the fractional $x$'s. This is indeed easy if spacetime is flat and the worldline of the observer is straight. The situation then is shown in fig. \ref{ff3}. If the observer has got a clock he is able to measure the time intervals between the arrivals of the pulses, i.e. the lengths between the light blue blobs marking the arrivals of the signals in fig. \ref{ff3}.

Under the geometrical assumptions we have made it is trivial to see that, given a sequence of eight arrivals, simple linear relations between the time intervals and the $x$'s hold. Solving the corresponding system of linear equations leads to the complete definition of the positions in spacetime of the first four reception events. With a moving sequence of events, it is then possible to fully reconstruct the whole worldline of the receiver. In order to turn our $n$'s and $x$'s into practical coordinates we must also know the proper emission periods of the sources (the $T$'s) and their positions in the sky, or, to say better, their worldlines, in practice the direction cosines of the propagation from each source (for details see \cite{Tartaglia2010,Tartaglia2010b}).

\section{Constraints and uncertainties}

In order to have an extremely simple algorithm we considered a flat spacetime and a straight worldline of the observer. Are these conditions credible? And to what extent?

\subsection{The observer's worldline}

In a flat spacetime a straight worldline corresponds to an inertial, then uniform, motion. This condition is not appropriate to describe the journey of a spacecraft in space both because of the presence of the gravitational field and of the maneuvers made using an engine; the more this is true if we consider the motion on the surface of a celestial body, including the Earth. In general the motion will be accelerated as in the case seen in fig. \ref{ff2}. Let us put, for the moment, the gravitational field aside. How can we treat the curvature of the worldline due to the acceleration?

Developing the worldline function in powers of time and looking at the first non linear term in the $i$-th space component of the motion we have of course:

\begin{equation}
s_i = v_it+\frac{1}{2}a_i t^2 \label{eq:para}
\end{equation}

 The relative importance of the non linear term with respect to the linear one is:

\begin{equation}
\epsilon = \frac{a_it}{2v_i} \label{eq:epsi}
\end{equation}

Now, if we decide what is the acceptable tolerance for our problem, we obtain the maximum time within which the linear approximation for the worldline (i.e. the uniform motion hypothesis) is viable:

\begin{equation}
t \leq \delta t = 2\epsilon \frac{v_i}{a_i} \label{eq:limite}
\end{equation}

Within the Solar System the speed of freely falling objects is controlled by Kepler's laws, so that it does not exceed  $\sim 10^5$ m/s (the escape velocity from the surface of the Sun is $\sim 6\times10^5$ m/s); for our purposes it is not the case to consider ``visitors'' not belonging to the Solar System. On the other hand for manned vehicles there are limitations of the maximal acceleration that should not exceed $30-40$ m/s$^2$ in order to prevent unacceptable physical damages (the acceleration at launch of the space shuttle is limited to approximately $30$ m/s$^2$). It is less simple to define upper limits to the tolerable acceleration for unmanned spacecrafts, since that limit would depend on the fragility of the payload and of the onboard equipments. It is however true that usually the preferred strategy for space missions tends to favor weak long-lasting thrusts rather than short and violent pushes, being the latter reserved at most for the take off from celestial bodies. Just to fix ideas we take as a reference highest acceleration the value of 100 m/s$^2$. Using these figures we see that an accuracy of 1 part in 10$^5$ implies that the linearity hypothesis is not tenable for more than approximately $\delta t = 10^{-2}$s. If we think to our sources as being millisecond pulsars, the above $\delta t$ includes a number of cycles which is roughly 10: this is enough for an 8 events sequence so that the actual worldline of the receiver is piecewise reconstructed as a chain of locally straight portions. Of course, if we think to sources with an emission period in the order of $\mu s$'s or less\footnote{In this case they could not be pulsars, since neutron stars cannot speed that fast, because of their size.}, we may use a much higher number of paces within a given interval. In this way, keeping the prescribed accuracy fixed, we could allow for much bigger accelerations.

The final accuracy of the positioning depends both on the quality and emission frequency of the sources and on the accuracy of the onboard clock used to measure the delay from one arrival event to the other.

\subsection{The gravitational field}

Until now we have neglected the influence of a gravitational field which is of course present throughout the Solar System. This means that the background spacetime is curved rather than flat; can we account for this? A first remark is that the curvature of spacetime at the emission point and along most of the trajectory of the signal, out of the Solar System, is irrelevant for our method, provided it does not change in time, or at least it changes only over times much much longer than the ones implied in our positioning process. The reason is that the curvature in the environment of the emitting neutron star and along the electromagnetic ray determines the global time of flight of a pulse in an (almost)
time-independent way, so that it does not affect the intervals between the arrival times of successive signals.
What remains to be considered is the effect of the curvature in the region where the receiver is located; since
the receiver is moving, also the local curvature it feels changes with time. However we remark that the gravitational
acceleration within the Solar System (excepting the surface of the Sun and the giant planets) is of the order
of 10 m/s$^2$ and usually much less than that. In these conditions a simple Newtonian approach is acceptable and 
the gravitational field can be treated as any other acceleration field in a flat background.
In practice for times short enough that the worldline can be considered to be straight we are on the tangent space and no gravitational field is visible, but for longer times gravity appears in the bending of the reconstructed spacetime
trajectory. Again we have a time tolerance within which our linear algorithm works; if we wish instead to use our method in order to investigate the structure of the gravitational field at the receiver's position, we simply need to collect data for a long enough time.

\section{An exercise}
\begin{figure}
	\centering
		\subfigure [ ] {
 		\includegraphics[width=1.05\linewidth]{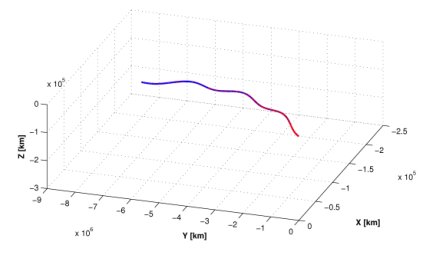}
		}
		\subfigure [ ] {
    \includegraphics[width=1.05\linewidth]{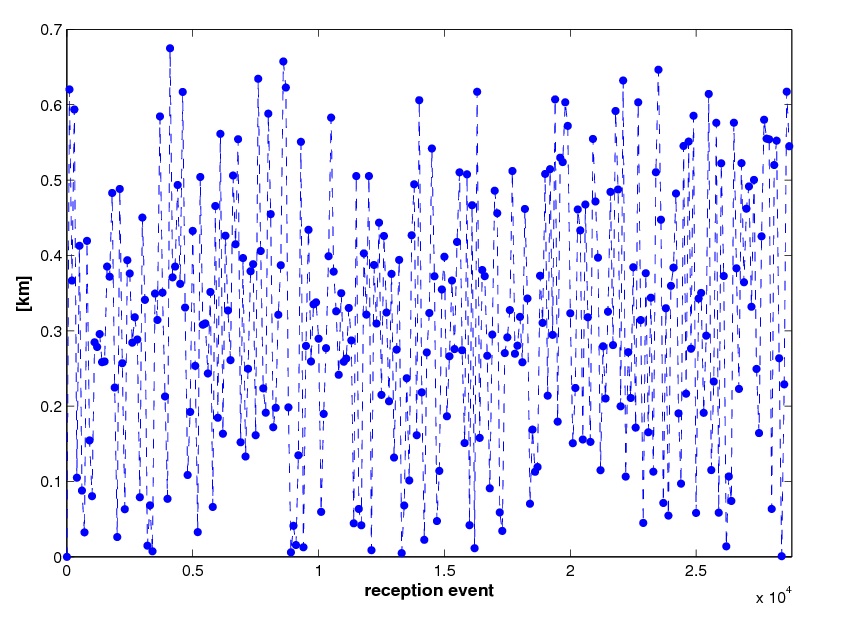}
		}
	\caption{Motion of the Earth with respect to the fixed stars during three days, reconstructed using the arrival times of four pulsars at the site of the Parkes observatory in Australia. (a) Global view; the scale does not allow to distinguish between the reconstructed trajectory and the ephemerides. (b) The values of the positioning error during the same three days; they are in the order of hundreds of meters.}
	\label{ff4}
\end{figure}

In order to make a preliminary evaluation of the feasibility of a positioning system like the one we have been describing so far, we have implemented our method developing a conversion algorithm from the arrival times measured by the receiver into the reconstructed spacetime evolution of the observer. Then we have applied our tool to the signals from four real pulsars as they would be received at the site of the Parkes observatory in Australia \cite{Ruggiero2010}. In practice the arrival times were simulated using the TEMPO2 programme, which has been developed by the astrophysics community to study the pulsars and can be used to simulate the sequence of the signals from known pulsars at any position on the surface of the Earth.
Our simulated data taking has been prolonged for three days and has produced the result synthetically visible in fig. \ref{ff4}.

In the figure one sees the space trajectory of the Earth with respect to the pulsars, assumed to be fixed stars. Actually the reconstructed orbit is superposed to the expected path given by the ephemerides. At the scale of the graph in a) it is impossible to catch any difference between the two curves, but in fact the simulated trajectory fluctuates about the fiducial one according to the assumed accuracy in the determination of the arrival times. The absolute positioning error during the same period and for about 30,000 points is visible in part b) of the figure; it is in the order of a few hundred meters.

\begin{figure}[ht]
\centering
\includegraphics[width=1.0\linewidth]{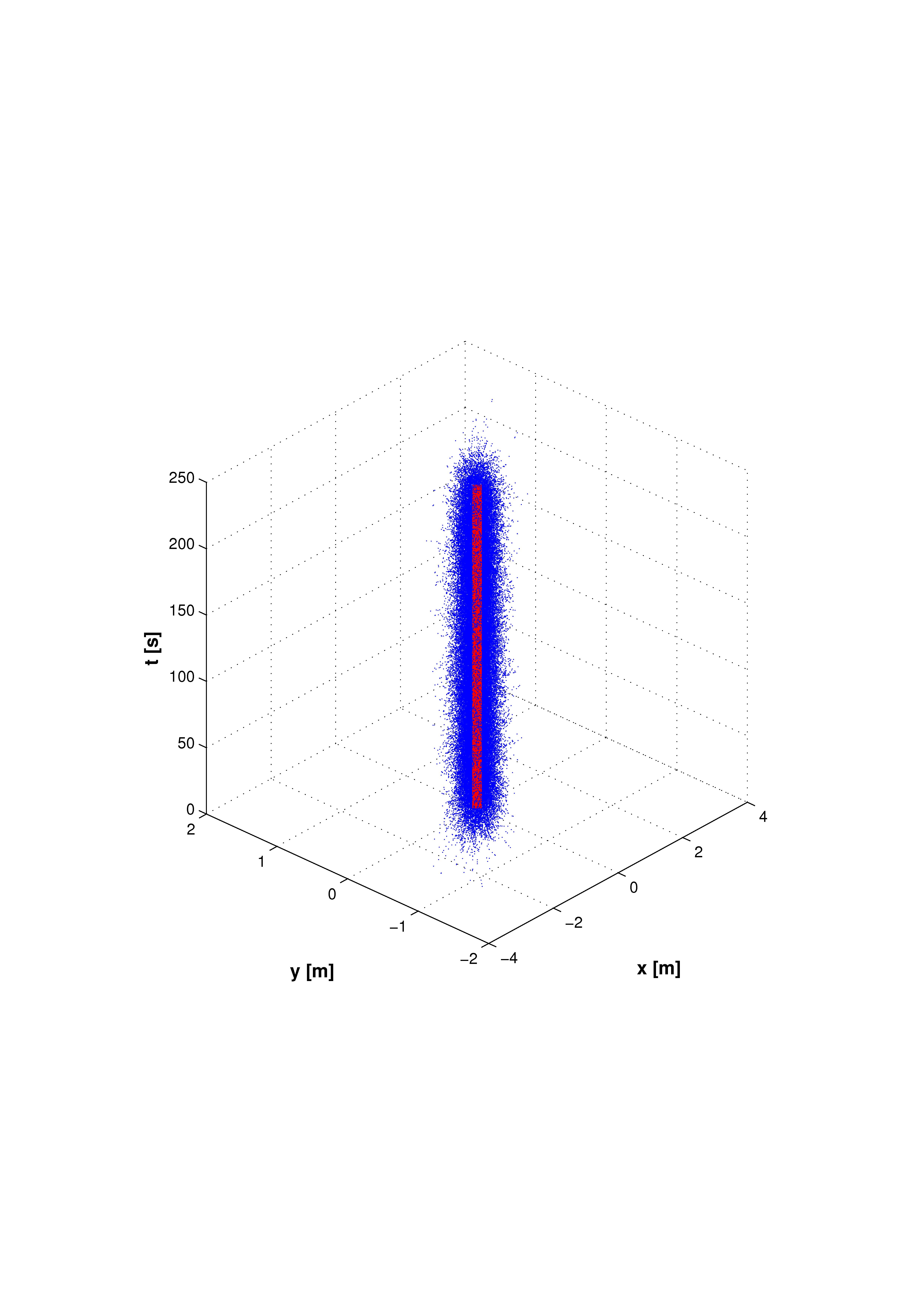}
\caption{Worldline of an observer at rest. The blue dots are obtained from the simulated arrival times of the signals
from four pulsars; the red line is the real spacetime position. The dispersion is contained into 40 cm.}
\label{ff5}
\end{figure}

In order to verify the response of our programme to the uncertainties in the measurement of time we have run some preliminary tests simulating an observer at rest with the fixed stars. A typical result (with one space dimension suppressed) can be seen in fig. \ref{ff5}. The asymmetry between the $x$ and $y$ directions is due to the non-fully symmetric distribution of the pulsars around the hypothetical user.

For the test we have simulated the actual behaviour of the onboard clock superposing to the fiducial arrival times a nanosecond Gaussian noise. Consistently the reconstructed position of the receiver displays a dispersion of the order of approximately 40 cm. In the exercise on the motion of the Earth, we prudentially adopted a few microseconds Gaussian noise.

\section{Fixed stars versus artificial sources}
The idea of using pulsars for navigation and positioning purposes was put forth since the early times of their discovery \cite{Downs1974} but everybody is perfectly aware of the fact that, though appealing, radio-pulsars are very faint objects so that their signals are many orders of magnitude below the noise at corresponding frequencies \cite{Sala2004}. This means that big antennas and refined elaboration techniques are needed in order to receive their pulses; the size of the antenna could be reduced considering X-ray pulsars, but the receiving device should be out of the atmosphere and the problem of the intensity of the signal remains \cite{Sheikh2006}. It is true that recognizing a known pulsar is not the same as looking for unidentified new objects, as people at radiotelescopes do, but anyway the use of pulsars would be difficult and limited to a few and expensive cases. Once however the method is adopted and validated there is no reason for excluding artificial pulsating sources. Actually if we allow for moving emitters we have an ample choice of different solutions. The relevant requisite is that the worldline of the emitter must be known; in practice in our algorithms we shall have time depending direction cosines and time depending relativistic $\gamma$ factors with respect to the reference frame in which we decided to represent both the sources and the user. The rate of change of the time depending values will have to be small enough not to spoil the linearization procedure for the receiver's worldline. If the emitter is an artificial one, we may of course think of using much higher frequencies than the ones we find in pulsars: GHz's will not be a problem. Since the piecewise linearization is made over times of a few tens of the emission periods at most, the relative stability will have to be kept over tens or hundreds of $\mu$s.

Suppose the source is carried by a spacecraft whose path in the Solar System is well known, or even by a satellite orbiting the Earth. The relative speeds of these vehicles with respect to the user can be of the order of, say, $\sim10^4$ m/s, which means that the emitter during the integration time of the algorithm moves at most 1 m. At a distance ranging from thousands to millions of km that displacement corresponds to angles from $\mu rad$ down to $nrad$ or less. Coming to the expected rate of change of the relative velocity of the emitter with respect to the receiver, we see that at the highest credible acceleration we have room for changes of the speed at most in the order of 1 cm/s. A change of this amount produces a corresponding change in the $\gamma$ in the order of 1 part in 10$^{19}$: perfectly negligible.

Similar results hold also when we imagine our emitters to be laid down on celestial bodies, such as the Moon, or Mars, or the asteroids.

Summing up, a permanent structure to enable the navigation in the Solar System could consist of a set of sources of electromagnetic pulses located on celestial bodies and on freely orbiting spacecrafts. Such a system could also be integrated by the use of a limited number of pulsars. In all cases, even though the minimum number of emitters is four, the sources on which to rely will have to be more than four: redundancy is important (as it is the case also for the present terrestrial GPS) because some of the sources, for periodic occultation or for any other reason, may turn out to be unavailable for a while. Let us add that for pure geometrical reasons the distribution of the sources in the sky of the user should always be as even as possible in order to optimize the accuracy of the positioning. This again implies a redundant number of emitters, especially if their position changes with time.

\section{Conclusion}
We have seen that it is possible, for positioning purposes, to exploit periodic pulses from sources with known worldlines. The use of null vectors makes the method intrinsically relativistic, so that no ad hoc correction is needed. If the time interval within which it is possible to treat the user's worldline as being straight is not less than ten times the longest period of the emitters, simple linear relations between the arrival times of the pulses from different sources hold. Solving the corresponding system of linear equations allows the reconstruction of the worldline of the user. Suppose you are starting a journey at a given moment and from a given position. You need to know: the position of the initial event with respect to an arbitrary reference frame; the position in the sky of at least four pulsating sources and their possible law of change with time; the period of the pulses from each source in a reference frame where the source is at rest.

Applying our algorithm the whole worldline of the traveler can be reconstructed with respect to the start event in the chosen reference frame. The accuracy of the positioning depends on the distribution of the sources in the various directions and on the precision of the clock the observer uses to measure the time from one pulse to the other.

The use of pulsars as sources has been considered, though it is limited by: the extreme weakness of the signals, the need for comparatively long integration times and the concentration of the pulsars in the galactic plane (uneven distribution in space). Anyway we have been able to simulate the reconstruction, by means of pulsars, of the worldline of a point of the surface of the Earth.

We have then discussed the application of our method to artificial emitters of pulses. In this case we would have the possibility to use higher frequencies, then shorter integration times, and much higher intensities, then a cheaper and more manageable hardware for the detection and treatment of the signals. In the case of an artificial source we need also to know the worldline of the emitter. In order to build a navigation support network in the Solar System we could think to place a number of our pulsating emitters on different celestial bodies and onboard freely falling spacecrafts.

Unlike the present situation, a space mission could be self-guided using the connection with the navigation support network. The spacecraft would have to be equipped with: an antenna apt to receive the pulses from the network and to recognize from which one of the sources they come; a clock; a memory containing the ephemerides of the sources from the initial event of the journey onward; a computing facility with a resident simple linear algorithm. No need for remote control would remain, except for some calibration from time to time. Of course the technological requirements of the equipment listed above have to be discussed carefully. For instance the antenna must be designed so that it can catch the pulses from at least four sources at a time; we would then need a set of either omnidirectional or wide aperture elements, with different orientation in space. Hopefully the determination of the direction from which the signals arrive is not necessary as far as the parameters of the source are entirely known, so that it will be sufficient to recognize the source from its ``signature''.

The attention has been concentrated on the space missions, however there is no reason to exclude more familiar applications for positioning on or around the Earth. Using a constellation of high Earth orbit satellites as primary sources would provide the equivalent of the present GPS, without the need for Sagnac periodic re-synchronization of the clocks from Earth. By the way, the next generation of Galileo satellites could be an opportunity to test our method in the terrestrial environment; for a practical test actually four micro-satellites in high Earth orbit broadcasting regular pulses could be enough.

The effort should now be concentrated on technology: clocks, receivers, processing units, energy consumption, miniaturization etc. The prospective of an autonomous navigation among the planets makes all this worth doing.

\bibliographystyle{plain}
\bibliography{actafutura_pulsarnv1}

\begin{thebibliography}{10}

\bibitem{Bini2008}
D.~Bini, A.~Geralico, M.L. Ruggiero, and A.~Tartaglia.
\newblock {Emission versus Fermi coordinates: applications to relativistic
  positioning systems}.
\newblock {\em Class. Quantum Grav.}, 25(20):205011, 2008.

\bibitem{Blagojevitc2002}
M.~Blagojevic, Garecki~J. F., W.~Hehl, and Yu.~N. Obukhov.
\newblock {Real null coframes in general relativity and GPS type coordinates}.
\newblock {\em Phys. Rev. D}, 65(4):044018, 2002.

\bibitem{Coll2006}
B.~Coll.
\newblock {Relativistic Positioning Systems}.
\newblock {\em AIP Conf. Proc.}, 841:277--284, 2006.

\bibitem{Coll2006b}
B.~Coll, J.~J. Ferrando, and Morales~J. A.
\newblock {Positioning with stationary emitters in a two-dimensional
  space-time}.
\newblock {\em Phys. Rev. D}, 74(10):104003, 2006.

\bibitem{Coll2006a}
B.~Coll, J.~J. Ferrando, and J.~A. Morales.
\newblock {Two-dimensional approach to relativistic positioning systems}.
\newblock {\em Phys. Rev. D}, 73(8):084017, 2006.

\bibitem{Downs1974}
G.S. Downs.
\newblock {Interplanetary navigation using pulsating radio sources}.
\newblock {\em NASA Technical Report}, (JPL-TR-32-1594, NASA-CR-140398):1--18,
  1974.

\bibitem{Rovelli2002}
C.~Rovelli.
\newblock {GPS observables in general relativity}.
\newblock {\em Phys. Rev. D}, 65(4):044017, 2002.

\bibitem{Ruggiero2010}
ML. Ruggiero, E.~Capolongo, and A.~Tartaglia.
\newblock {Pulsars as celestial beacons to detect the motion of the Earth}.
\newblock {\em Int. J. Mod. Phys. D}, In press:10, 2010.

\bibitem{Ruggiero2008}
M.L. Ruggiero and A.~Tartaglia.
\newblock {Mapping Cartesian coordinates into emission coordinates: Some toy
  models}.
\newblock {\em Int. J. Mod. Phys. D}, 17(2):311--326, 2008.

\bibitem{Sala2004}
J~et~al. Sala.
\newblock {Feasibility Study for Spacecraft Navigation System Relying on Pulsar
  Timing Information}.
\newblock {\em European Space Agency, the Advanced Concepts Team, Ariadna Final
  Report}, (03-4202), 2004.

\bibitem{Sheikh2006}
S.~I. et~al. Sheikh.
\newblock {Spacecraft Navigation Using X-Ray Pulsars}.
\newblock {\em J. Guid. Control Dynam.}, 29(1):49--63, 2006.

\bibitem{Tartaglia2010}
A~Tartaglia.
\newblock {Emission coordinates for the navigation in space.}
\newblock {\em Acta Astronaut.}, 67:539--545, 2010.

\bibitem{Tartaglia2010b}
A.~Tartaglia, ML. Ruggiero, and E.~Capolongo.
\newblock {A null frame for spacetime positioning by means of pulsating
  sources}.
\newblock {\em Advances in Space Research}, 47:645--653, 2011.

\end{thebibliography}

\end{document}